# Online External Beam Radiation Treatment Simulator

Felix G. Hamza-Lup[a], Ivan Sopin[a], Omar Zeidan[b]

[a] Computer Science, Armstrong Atlantic State University, Savannah, Georgia, USA
[b] M.D. Anderson Cancer Center Orlando, Orlando, Florida, USA

**Abstract.** Radiation therapy is an effective and widely accepted form of treatment for many types of cancer that requires extensive computerized planning. Unfortunately, current treatment planning systems have limited or no visual aid that combines patient volumetric models extracted from patient-specific CT data with the treatment device geometry in a 3D interactive simulation. We illustrate the potential of 3D simulation in radiation therapy with a web-based interactive system that combines novel standards and technologies. We discuss related research efforts in this area and present in detail several components of the simulator. An objective assessment of the accuracy of the simulator and a usability study prove the potential of such a system for simulation and training.

*Keywords:* external beam radiation therapy, distributed training, three-dimensional simulation

## 1. Introduction

External beam radiation therapy (EBRT) is an effective and widely accepted form of treatment for many types of cancer that requires extensive planning. Unfortunately, current treatment planning systems have limited or no visual aid that combines patient volumetric models extracted from patient-specific computed tomography (CT) data with the treatment device in an inclusive 3D interactive simulation. Therefore, treatment planners often find it difficult to determine precise treatment equipment setup parameters. In some cases, patient treatment is delayed or postponed due to unforeseen collisions. In addition, the demand for better cancer targeting has created specific immobilization and on-board imaging devices, which can become additional collision sources. We present a web-based EBRT simulation system developed through a multidisciplinary research effort. We illustrate the system's user interface, and interaction paradigms as well as the technologies used. We also provide a comprehensive simulation accuracy assessment.

The paper is structured as follows. Section 2 presents the simulator's scope and recent work by other research and development groups. In section 3, we illustrate the main components of the web-based system, emphasizing the user interaction and we provide the details of the polygonal models acquisition process. In section 4, we discuss simulation assessment strategies and current results. We conclude, in section 5, with a discussion on the importance of online simulation and training systems and new technologies for the medical field.

## 2. Rationale

In the planning process, radiation therapy planning systems make use of 3D visualization of volumetric data. This data usually describes radiation doses which have to be delivered at specific locations to destroy cancerous tissue. Since the success of the treatment depends on the accuracy of planning and delivery, physicians need robust tools to assist them in the planning process. For the past fifteen years, numerous software modules have been added to the planning systems; however, none of them provide adequate collision detection (CD) and room level setup visualization.

A wide range of analytical methods for linac (linear accelerator)-based radiation therapy have been proposed in the past as means to improve the EBRT planning process [1-4]. Some of these methods, even though accurate, are based on hardware numeric rotational and translational values, disregarding patient-specific and detailed hardware-specific geometry. Previous research and development concerning graphical simulations of linac systems have limitations such as:

- Simulations involve only generic patient body representations [5] and not accurate hardware 3D models; hence, collisions with patients are not accurately modeled and predicted;
- Simulations run as standalone applications and cannot be deployed over the web for potential collaboration with remote experts during treatment planning.

One of the graphical simulations of the patient and hardware is under development at Hull University [6]; however, the current implementation cannot be deployed freely over the Internet and does not have a precise CD tool. Our efforts, directed towards the development of a comprehensive yet simple to deploy web-based simulator, generated a first prototype in 2005. Since then, we have improved the system to provide an accurate representation of the EBRT room setup based on patient-specific medical data. We deployed the latest version of the system on a secure website (www.3drtt.org) in 2007 and provided free registration for interested parties. More than one hundred users have registered and used the simulator. We have recently started recording visitors' geographical locations based on their IPs, with the ability to display them on a map using the Google Maps public API. According to our 2007 records, cancer treatment professionals from more than twenty countries around the world have visited the website.

## 3. Simulator Description

### *3.1 Methods and Tools*

Our simulator implementation takes advantage of the X3D standard [7] and several scripting languages (i.e., JavaScript, AJAX, and JSP); in the development process, we also employ several software tools for 3D modeling. X3D is an ISO standard with an open architecture and a rich range of capabilities for real-time graphics processing. A successor to VRML, X3D is being developed by the Web3D Consortium as a more mature and refined standard. Fig.1 illustrates snapshots of the web-based simulator (denoted 3D Radiation Therapy Training — 3DRTT), which provides a room-based view of the hardware and patient setup.

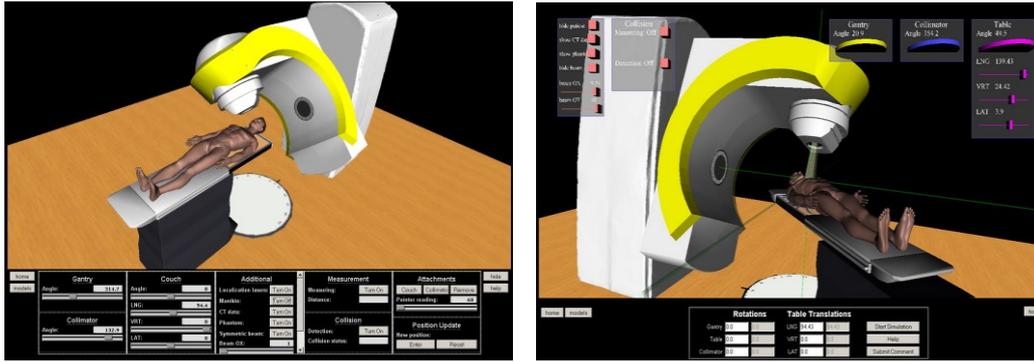

Fig. 1. 3DRTT with HTML/JavaScript-based (left) and X3D-based (right) graphical user interface

*3.2 Online Simulator Performance*

At initialization (i.e., first web access) all graphical components are downloaded as X3D files from the server. Depending on the X3D client's cache settings, a significant part or all of the files can be cached locally and used on the next access session if no updates are found on the server, significantly reducing network traffic. All of the scene interactions are computed locally, improving the system's scalability. Multiple users accessing the simulator at the same time experience no concurrency delays.

The 3D scene contains approximately 100,000 polygons in the basic setup (without additional components such as head attachments, phantom device, etc.). On all of the computers we tested, the graphics were rendered at high frame rates (e.g., 30 FPS or more). The only noticeable drop in frame-rate observed occurs when the CD mode (see section 3.3.4) is activated to register collisions with large objects. The problem can be solved by reducing the number of polygons in the CT data representation. In the current implementation detecting gantry/couch and collimator/couch collisions does not visibly affect the performance of the simulator.

*3.3 Graphical User Interface*

In the following sections, we describe several user interface components of the simulator, enabling measurements, hardware configurations, and immobilization attachments.

*3.3.1 Menus and Controls*

3DRTT explores new web-enabled graphic user interfaces. We provide two GUI versions: X3D-based and HTML/JavaScript-based. The X3D-based GUI (Fig. 2) enables the user to control and navigate through the virtual environment via the built-in X3D player features and a collection of movable semitransparent volumetric menus [8]. Each menu holds specific GUI components that refer to a particular object or function of the simulator (e.g., gantry, couch, radiation beam, etc.). Such 3D elements naturally mimic the behavior of the controlled objects or features: sliders for translation, scrolls for rotation, and buttons for switching between different modes.

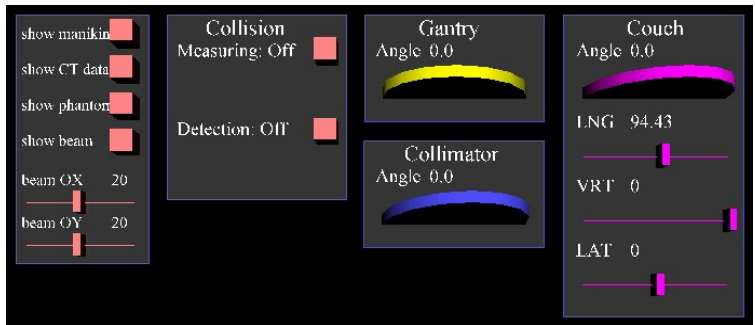
Fig. 2. X3D-based GUI

In the HTML/JavaScript-based version (Fig. 3), instead of custom-made volumetric menus, familiar Windows-like controls, such as buttons, scrolls, and text inputs, are provided. This flexibility enhances the usability of the simulator (e.g., the interface learnability, specifically familiarity).

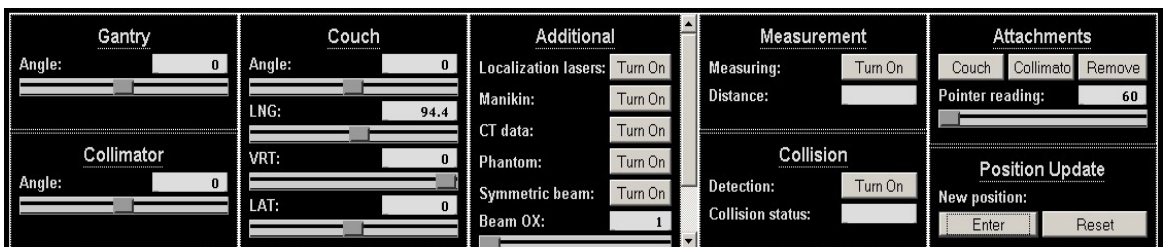
Fig. 3. HTML/JavaScript-based GUI

### 3.3.2 Virtual Measurement Tool

The measurement tool (Fig. 4) allows users to determine the exact distance between any two points in the virtual space. Users may control the locations of two small spheres by changing their Cartesian coordinates. The spheres are interconnected by a line, and the current distance between them is continuously displayed.

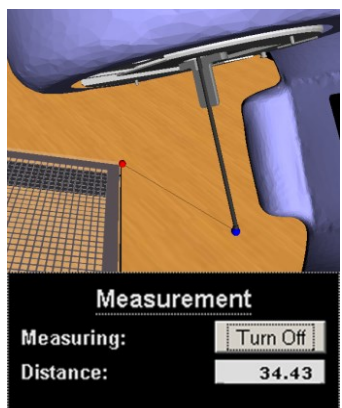
Fig. 4. Measurement mode

The tool enables CD validation and simulator accuracy assessment. In a near-collision scenario it is important to know the distance between two potential collidees such that the treatment plan can be modified accordingly. In the assessment process our next step is to determine the accuracy of the simulator in conjunction with the real environment.

To further improve the experience with spatial estimations, certain 3D visualization devices (e.g., 3D shutter glasses or other 3D displays) could be easily employed.

### 3.3.3 Selecting from an Interactive 3D Menu

Radiation treatment procedures require the use of additional accessory equipment for patient-specific treatments such stereotactic cones, electron cones, or blocks that attach to the collimator. In addition, various types of immobilization and positioning devices like stereotactic head frames and breast boards are attached to treatment couch. These accessories add to the complexity of planning process as additional precautions need to be accounted for to compute the clearances of a certain plan setting when such components come into play.

Our simulator provides an intuitive way to bring this equipment into the scene (Fig. 5). Each object can be loaded and removed individually.

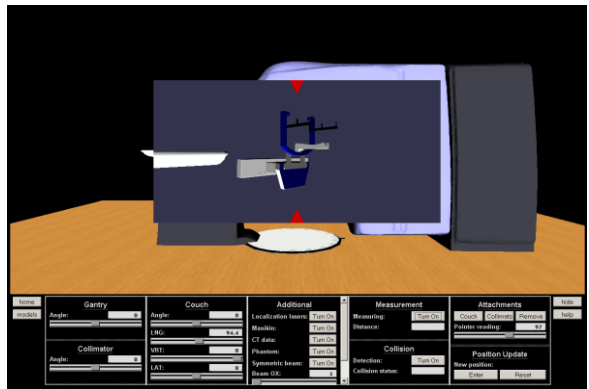

Fig. 5. Attachments menu

### 3.3.4 Collision Detection

Our preliminary web-based CD algorithm was based on the approximation of colliding objects with geometric primitives, as illustrated in Fig. 6. A tradeoff between accuracy and performance was considered.

Recently, an implementation of polygonal level CD became available in several X3D players. The *computeCollision()* function supported by the Bitmanagement™ Contact X3D player has four arguments: the source and target geometry nodes and their matrices of transformation. The target geometry node can be a composite grouping object that incorporates other geometry subnodes, but the source node should point directly to the object's geometry and cannot include other nodes. For this reason, we use source objects that are solid and do not contain nested subnodes.

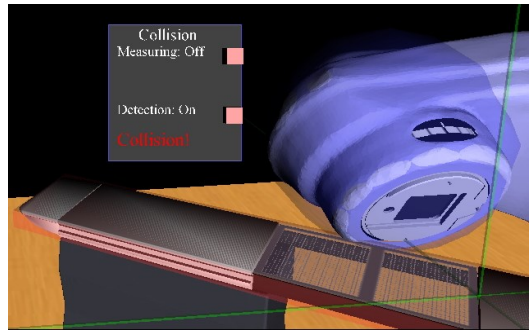
Fig. 6. CD based on Bounding objects

Our primary application for the *computeCollision()* function is to detect collisions between the couch and the gantry. Several techniques are used to speed up the CD processing. For instance, we represent the geometry of the couch with a parallelepiped, which approximates the real shape very well. Because of the gantry's complex geometry, we apply the product of the parallelepiped's and gantry's transformation matrices to place the "couch" into the gantry's coordinate system. In this way, we eliminate excessive calculations for gantry transformations and speed up the computation process.

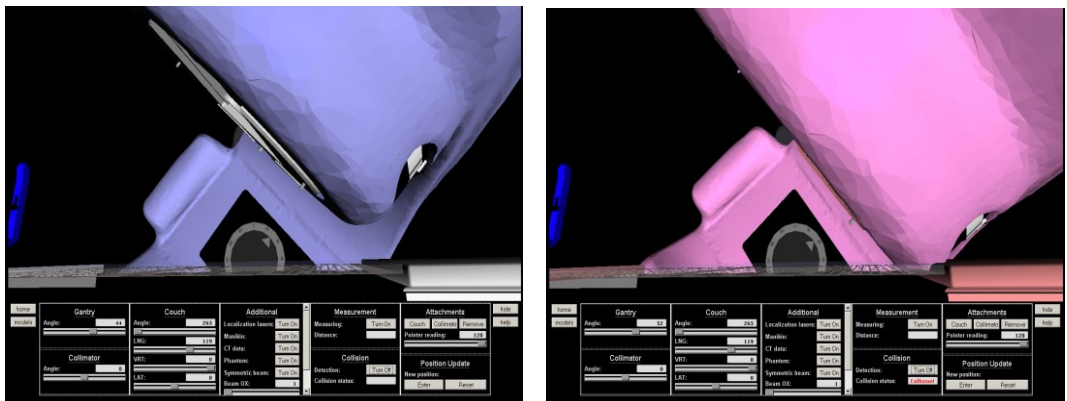
Fig. 7. Visualization of collision-free (left) and collision (right) scenarios.

Once the CD system is activated, it provides information about collisions in the hardware setups by highlighting the collidees and displaying a collision-warning message in the control panel (Fig. 7). If any couch attachments are currently installed, they are automatically included in the CD. Collision detection for embedded patient-specific CT data has been successfully tested using the current method and is currently under assessment.

### 3.4 Polygonal Models Acquisition

### 3.4.1 3D Scanning and Optimization

We use Faro™ LS 840 and the Konica-Minolta VIVID-9i 3D laser scanners to obtain collections of point clouds for Varian 23iX Trilogy® and Novalis®. We have found the VIVID-9i scanner to be the best solution. Several point clouds are merged into a single cloud for each model based on the location of special markers that are positioned in the treatment room throughout the scanning. Next, we filter the noise from the point cloud and wrap the rest into a polygonal surface. The size of the polygonal model must be further optimized for the Web environment. Further optimizations consist in removing redundant polygons in the areas with the least curvature. All wrapping and polygonal processing is done with Geomagic® Studio algorithms. The X3D standard allows for further optimizations of the scene, improving both realism and rendering efficiency.

*3.4.2 Manual Measurements*

Laser scanners are powerful tools in obtaining precise virtual 3D representations of physical objects. However, some objects possess surface properties that render laser scanning useless. For reflective surfaces correct registration of laser beams is difficult. In such scenarios manual measurements can provide better results. For the 3DRTT simulator, we manually record (using digital calipers) the dimensions of various couch and collimator attachments and use them to accurately reproduce those components in 3D at sub-millimeter accuracy. The use of digital photography (at orthogonal viewpoints) is also instrumental in modeling and validating the shapes. We use SolidWorks™ and 3D Studio Max™ software products for modeling these components.

*3.5 Patient 3D Model Generation from DICOM-RT CT*

An important issue we address in this project is the inclusion of real patient data in the simulation. We convert a patient's set of CT scans into a volumetric representation of the boundaries (i.e., the shell, Fig. 8). The set of CT scans is usually stored using DICOM RT standard [9]. As opposed to other types of image files, the DICOM RT files contain slice resolution, slice spacing, and pixel size, which are useful parameters in producing a realistic polygonal model of a patient. We process the CT scans using algorithms similar to the ones in the Visualization Toolkit (VTK) [10], with additional optimizations for rendering speed and accuracy. We apply the Marching Cubes algorithm with a value that selects the isosurface of the patient's skin (Fig. 8, middle). Then, the 3D model is converted into X3D and embedded in the simulator (Fig. 8, right).

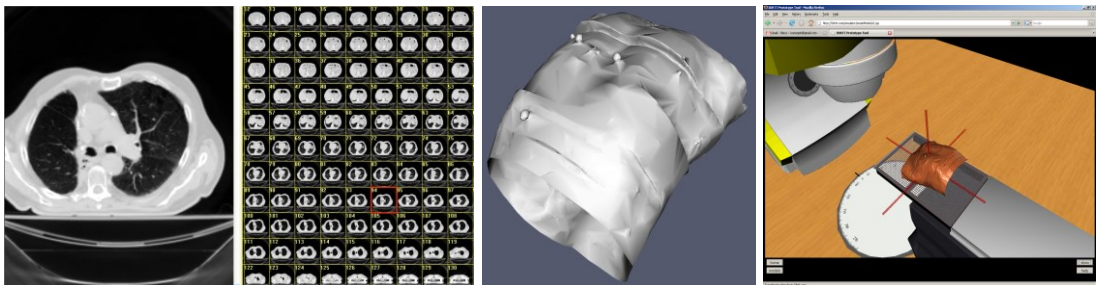

Fig. 8 CT data, 3D model reconstruction and 3D model inside simulator

For calibration, and sometimes training,, various predefined shapes are used by the medical personnel. An example is the elliptical phantom device (Fig. 9), used for testing and training. Such phantoms usually have a regular shape that is known beforehand. We are in the process of developing a library of 3D models that will augment the existing simulation with similar components.

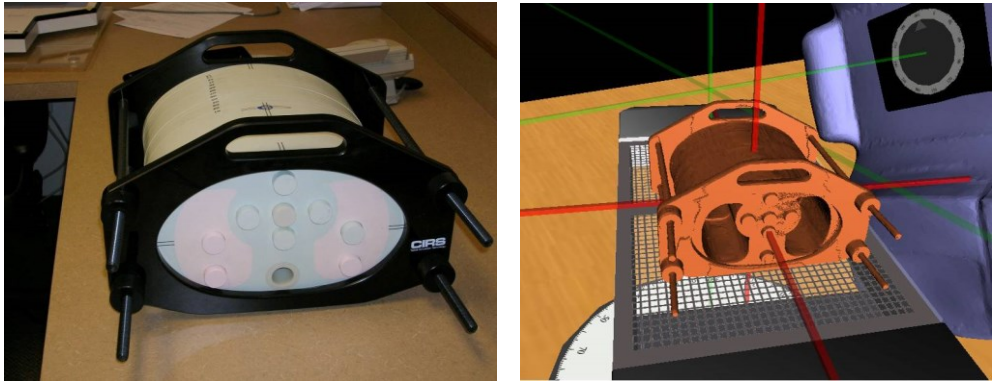
Fig. 9 The phantom device: real environment (left) and simulated environment (right)

## 4. Simulator Accuracy Assessment

Accuracy assessment objectively indicates how closely the simulated geometry corresponds to the real hardware and how well the simulator matches the actual clinical setups. Since we cannot obtain the deviation distribution for the entire reproduced geometry of the linac, one solution is to run identical real and virtual (simulated) treatment scenarios and compare the clearance among the hardware components in near-collision cases.

The measurements were done at M.D. Anderson Cancer Center Orlando. We ran twenty collision and near-collision scenarios, measuring (with digital calipers) the distances between the (potential) collidees. We reproduced these scenarios in our virtual setting and measured the same distances using the virtual measurement tool (see section 3.3.2). We focused on collimator/couch and collimator/head-fixation attachment interactions, as they are a frequent cause of collision in reality [3]. Fig. 10 provides visual comparisons of collision scenarios between the Varian® and Novalis® linacs and our corresponding simulators.

| LINAC Model | Real Machine | 3DRTT Simulator |
|---|---|---|
| Varian | 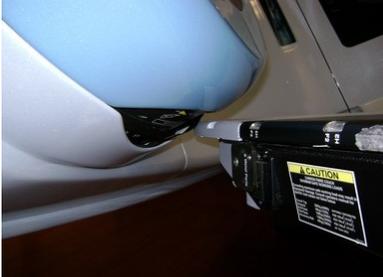 | 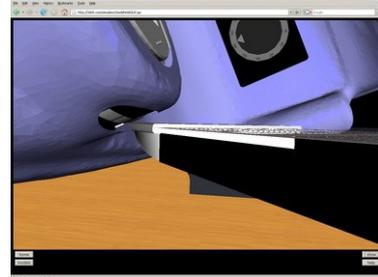 |
| Varian | 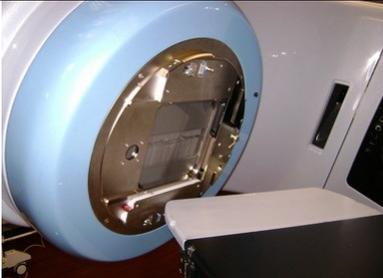 | 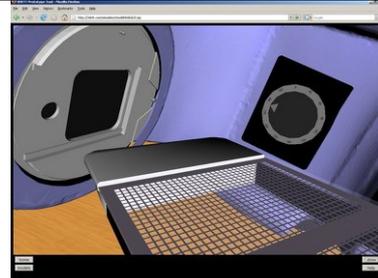 |

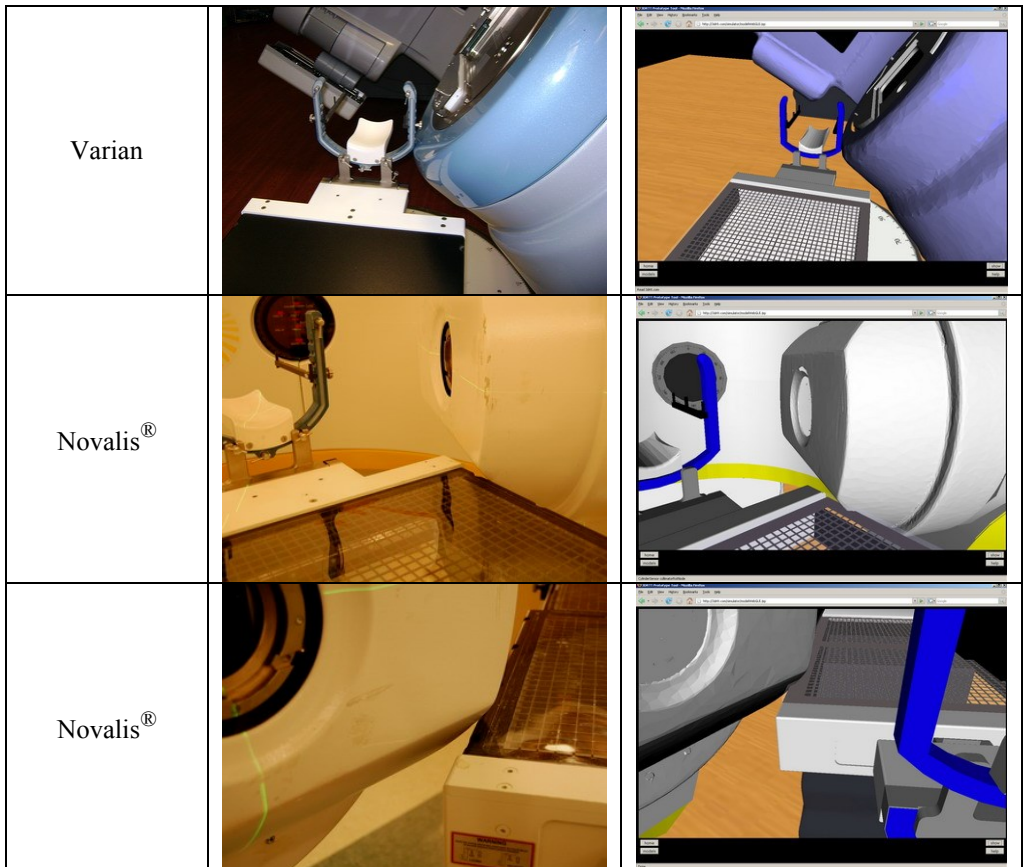

Fig. 10 Illustration of real (left) and simulated (right) collision scenarios

The objective assessment resulted in a mean difference of 1 cm between the Novalis® simulator and the real setup, with a standard deviation of 0.57 cm. The measurements were taken with a head-fixation attachment which acted as a (potential) collidee in half of the cases.

The Varian simulator yielded better accuracy with a mean difference of 0.5 cm. As illustrated in Fig. 10, the Varian Trilogy™ and the virtual representation look identical in the collision scenarios. Geometries of gantry and collimator for this simulator were modeled based on point clouds collected with the Faro™ Technologies laser scanner (see section 3.4.1). This technique contributed to the improvement in accuracy as compared to the Novalis® model, which was built using a few manual measurements (with digital calipers).

Radiation oncologists who collaborated with us on the 3DRTT project consider that an overall 1 cm accuracy is sufficient and will help significantly in obtaining a visual representation of the treatment equipment and patient setup improving the confidence in the plan delivery. Ultimately, the simulator's accuracy can be further enhanced by better polygonal model acquisition techniques (e.g., use of higher-resolution laser scanners), which will reduce geometrical errors to negligible level.

## 5. Conclusions

We have presented an online EBRT simulator that helps detect possible collisions between linac components (including machine add-on accessories and immobilization devices) for a given patient. The simulator may eliminate the need for backup plans and may save patients' treatment time and money.

Benefits of using the online EBRT simulator are educational and clinical. For the education component, the targeted trainee groups include physics residents, dosimetry trainees, and radiation therapy technicians. The online web-based system will allow trainees and students to
- Familiarize themselves with the translational and rotational motion limits of various linac components, including couch, gantry, and collimator, and with their angle conventions.
- Validate patient setups and plan deliverability and check for possible collision scenarios and beam-couch intersections;
- Educate patients about their treatment delivery technique and help reduce pre-treatment anxiety.

In addition to its educational benefits, the web-based system can be used clinically to improve the overall quality of radiation therapy, enabling the planner to see and better explore differing and unconventional gantry-couch-collimator combinations that may improve the treatment efficiency.

The main advantages of our system, as compared with other systems are, remote distributed access (e.g., providing expert advice distantly); 3D patient-specific data merged into the scene; precise modeling and accurate collision detection; and a thin low-cost software client that is easy to deploy in a hospital setup. Additionally, our approach to 3D simulation for CD and simulation is generic and may be adapted for other equipment or clinical settings. The methodology can be extended and used for all types of radiation therapy devices. The system has the potential for education: not only for medical students, but also for patients.

**Acknowledgements**

We would like to thank the Radiological Society of North America for the Education Seed Grant, Grant ID ESD0705, RSNA R&E Foundation as well as M.D Anderson Cancer Foundation for the support provided for this project.